\documentclass[%
reprint,
superscriptaddress,
amsmath,amssymb,
aps,
prl,
floatfix,
]{revtex4-1}

\usepackage{graphicx}
\usepackage{amsmath}
\usepackage{amssymb}
\usepackage{slashed}
\usepackage{bbm,bm}
\usepackage{hyperref}
\usepackage{xcolor}
\usepackage{comment}
\usepackage{amsfonts}
\usepackage{dsfont}
\usepackage{soul}
\usepackage{multirow}
\usepackage{placeins}

\def\bea{\begin{eqnarray}} 
\def\eea{\end{eqnarray}}
\def\be{\begin{equation}}
\def\ee{\end{equation}} 
\def\ba{\begin{array}}
\def\ea{\end{array}}

\newcommand{\avg}[1]{\left< #1 \right>}

\newcommand{\eps}[0]{\varepsilon}

\newcommand{\vx}{\vec{x}}
\newcommand{\vy}{\vec{y}}
\newcommand{\vp}{\vec{p}}

\usepackage{xcolor}

 \usepackage[normalem]{ulem}

\begin{document}
%%%%%%%%%%%%%%%%%%%%%%%%%%%%%%%%%%%%%%%%%%%%%%%%%%%

\title{Eulerian spatio-temporal correlations in passive scalar turbulence}

\author{Anastasiia Gorbunova}
%\email{anastasiia.gorbunova@univ-grenoble-alpes.fr}
\affiliation{University Grenoble Alpes, CNRS, LPMMC, 38000 Grenoble, France}
\affiliation{University Grenoble Alpes, CNRS, Grenoble INP, LEGI, 38000 Grenoble, France}

\author{Carlo Pagani}
\affiliation{University Grenoble Alpes, CNRS, LPMMC, 38000 Grenoble, France}

\author{Guilaume Balarac}
\affiliation{University Grenoble Alpes, CNRS, Grenoble INP, LEGI, 38000 Grenoble, France}
\affiliation{Institut Universitaire de France, 1 rue Descartes, 75000 Paris, France}

\author{L\'eonie Canet}
\email{leonie.canet@grenoble.cnrs.fr}
\affiliation{University Grenoble Alpes, CNRS, LPMMC, 38000 Grenoble, France}
\affiliation{Institut Universitaire de France, 1 rue Descartes, 75000 Paris, France}

\author{Vincent Rossetto}
\affiliation{University Grenoble Alpes, CNRS, LPMMC, 38000 Grenoble, France}

%%%%%%%%%%%%%%%
\begin{abstract}
%%%%%%%%%%%%%%%
We study the spatio-temporal two-point correlation function of passively advected scalar fields
in the inertial-convective range in three dimensions by means of numerical simulations.
 We show that at small time delays $t$ the correlations decay as a Gaussian in the variable $tp$ where $p$ is the wavenumber.
 At large time delays, a crossover to an exponential decay in $tp^2$ is expected 
  from a recent functional renormalization group (FRG) analysis. 
  We study this regime for a scalar field
   advected by a Kraichnan's ``synthetic'' velocity field,
    and accurately confirm the FRG result, including the form of the 
    prefactor in the exponential. By introducing finite time correlations
     in the synthetic velocity field, we uncover the crossover between the two regimes.
%%%%%%%%%%%%%%%%
\end{abstract}
%%%%%%%%%%%%%%%%

\maketitle

The advection of a scalar field by a turbulent flow plays a crucial role 
in  several domains ranging from engineering to geophysics. 
For instance, a scalar field can represent the concentration of a chemically inactive impurity, 
or a temperature fluctuation. It is coined as passive when
 its backreaction on the flow is negligible.
We focus on passive scalar fields
 in the inertial-convective range, which spans between the large scale at which the energy is injected and the small scale at which the dissipation occurs.
In this range, the turbulent flow and the advected scalar both
 conform to the Richardson's cascade picture, characterized by a constant downscale energy flux.
In  seminal works, Obukhov and Corrsin \cite{obukhovStructureTemperatureField1949,corrsinSpectrumIsotropicTemperature1951} established that the 3D energy spectrum 
 of the scalar field 
in this range decays as $E_\theta \left(\vp\right)\sim p^{-5/3}$, hence exhibiting the same scaling as the one predicted by Kolmogorov in his 1941 statistical theory of turbulence \cite{Kolmogorov41,Kolmogorov41a}.
\par  Despite much understanding has been gained 
 on the statistical properties of scalars in fluid flows
\cite{sreenivasanTurbulentMixingPerspective2019, warhaftPassiveScalarsTurbulent2000, shraimanScalarTurbulence2000,falkovichParticlesFieldsFluid2001},
certain aspects remain elusive.
 In particular, the time dependence of Eulerian correlation functions of a passive scalar field 
 in a turbulent steady state is poorly understood. 
 Its comprehension is essential for various applications, such as 
 the development of time-accurate numerical models \cite{heSpaceTimeCorrelationsDynamic2017}, experimental data treatment \cite{heKraichnanRandomSweeping2011}
and turbulent diffusion problems \cite{majdaSimplifiedModelsTurbulent1999}.
 In early works \cite{kraichnanKolmogorovHypothesesEulerian1964, tennekesEulerianLagrangianTime1975}, it was suggested that the temporal properties of
Eulerian correlation functions are determined by the random sweeping effect, i.e., 
the advection of small-scale velocities by random large-scale motion, which leads to a  scaling of
 the correlation time as $p^{-1}$.
Although the dominance of the sweeping effect has been discussed 
 on phenomenological bases and observed in numerical simulations \cite{chenSweepingDecorrelationIsotropic1989, yeungRandomsweepingHypothesisPassive2002, ogormanModalTimeCorrelations2004} and experiments \cite{heKraichnanRandomSweeping2011}, 
 a more rigorous theoretical justification was missing.
  
\par  Significant advances were provided by the analysis of 
 simplified models of scalar turbulence, such as the model proposed by Kraichnan  \cite{kraichnanSmallScaleStructureScalar1968},
 in which the Navier-Stokes (NS) velocity field is replaced by a random vector field with a white-in-time Gaussian statistics.  
In suitable limits,
this simplification indeed allows for explicit analytical calculations of the anomalous scaling exponents of the structure functions
via different approaches 
\cite{ChertkovFalkovich1996,ChertkovETal1995,Gawedzki:1995zz,Bernard:1996um,BernardGK1998,Adzhemyan1998,Adzhemyan2001,Kupiainen:2006em,Pagani:2015hna},
 we refer to \cite{shraimanScalarTurbulence2000,falkovichParticlesFieldsFluid2001, antonovRenormalizationGroupOperator2006} for reviews.
 Furthermore, the temporal dependence of the scalar correlation function was analyzed in \cite{mitraDynamicsPassiveScalarTurbulence2005, sankarrayUniversalityDynamicMultiscaling2008}.
  However, the simplified velocity covariance prevents one from relating the model to real scalar turbulence. 

  Recently, the study of the temporal properties of correlation functions in a turbulent steady-state
  has received a boost triggered by the use of functional
   renormalization group (FRG)
 \cite{dupuisNonperturbativeFunctionalRenormalization2021}.
 For NS turbulence, 
  this framework allows one to obtain an analytical expression
   for arbitrary correlation functions in  the limit of large wavenumbers.
This arises from the fact that the FRG flow equations
 for these objects can be closed in a large wavenumber expansion
    exploiting  only  the symmetries and the related Ward identities \cite{canetSpatiotemporalVelocityvelocityCorrelation2017,tarpinBreakingScaleInvariance2018,tarpinStationaryIsotropicHomogeneous2019}.
 Remarkably, within the FRG framework, 
 one can derive an approximated form of the 
spatio-temporal Eulerian two-point correlation function of the scalar field advected by NS flows
 and in Kraichnan's model \cite{paganiSpatiotemporalCorrelationFunctions2021}. 
 
In this Letter, we study scalar fields 
advected by NS flows as well as synthetic velocity fields in three dimensions by means of direct numerical simulations (DNS).
 We focus on the spatio-temporal dependence of the Eulerian
 two-point correlation function of the scalar field.
 We find
  that it exhibits two distinct
   time regimes in the stationary state:
   a Gaussian decay in the variable $tp$ at small time differences $t$
    followed by an exponential decay in $tp^2$ at large time differences,
    although the latter is only evidenced for the synthetic velocity field.
     Beyond the general form of the correlations, we also compute the non-universal prefactors
     in the exponentials,
    and show that these results are in accurate
      agreement with the FRG predictions, thereby  providing
       a detailed and precise account of the Eulerian temporal behavior of  scalar turbulence.

\paragraph{Dynamics of the passive scalar field.}

We consider a scalar field $\theta(t,\vx)$ governed
  by the following advection-diffusion equation
\begin{equation}
\partial_t \theta +v^j \partial_j \theta =
\kappa \partial^2 \theta +f_\theta\,, \label{eq:scalar-dyn-1}
\end{equation}  
where $\kappa$ is the molecular diffusivity and $f_\theta$ a stochastic forcing
  peaked at some large scale $L$.
The advection  field $\vec{v}(t,\vx)$ is the velocity of an incompressible fluid satisfying
 the {NS} equation,
\begin{equation}
\partial_t v^i +v^j \partial_j v^i =
-\frac{1}{\rho} \partial_i \pi + \nu \partial^2 v^i + f^i\,,
\label{eq:NS-eq}
\end{equation}
where {$\pi$} denotes the pressure, $\rho$ the density, $\nu$ the viscosity,
and $f^i$ a stochastic forcing  
which injects energy at the same integral scale $L$. 
 The scalar is assumed to be passive, which means that it does not
  affect the carrier fluid.

We also consider  Kraichnan's model~\cite{kraichnanSmallScaleStructureScalar1968},
 in which the velocity is a random Gaussian vector field,
   characterized by zero mean and covariance
\begin{equation}
\bigr\langle v^{i} \left(t,\vx\right) v^{j} \left(t^{\prime},\vy \right) \bigr\rangle  
= 
\delta\left(t-t^{\prime}\right) D_{0}\int_{\vp} \frac{e^{i \vp\cdot\left(\vx-\vy\right)}P_{ij}(\vp)}{\left(p^{2}+m^{2}\right)^{\frac{d}{2}+\frac{\varepsilon}{2}}}
 \label{eq:vel-2pt-correlation-function} 
\end{equation}
where $P_{ij}(\vp)\equiv \delta_{ij}-\frac{p_i p_j}{p^2}$
 is the transverse projector which ensures incompressibility
 and $\int_{\vp} \equiv \int d^d \vp $.
The parameter $ 0<\varepsilon/2<1$ corresponds to the H{\"o}lder exponent, 
describing the velocity roughness from very rough for $\varepsilon\to 0$ to smooth for $\varepsilon\to 2$, 
 and $m$ acts as an IR cutoff. 

\paragraph{Theoretical results from FRG.}

The FRG is a modern and versatile implementation of the Wilsonian RG \cite{dupuisNonperturbativeFunctionalRenormalization2021}.
Within this formalism, a generic $n$-point correlation function is calculated by
 deriving an equation governing its dependence on the RG scale, and solving  it 
   via some approximation scheme.
  For scalar turbulence, it turns out that the flow equation 
   of the two-point function can be closed for large wavenumbers, 
  thanks to the symmetries of the field theory associated to Eqs.~(\ref{eq:scalar-dyn-1})
and (\ref{eq:NS-eq})  \cite{paganiSpatiotemporalCorrelationFunctions2021}.
No additional approximation is needed besides the large wavenumber limit, 
 in particular these results do not rely on any small expansion parameter,
  and this closure can be achieved for scalars advected by NS flows, as well as for Kraichnan's model.

   In the case of NS flows, the FRG yields for the two-point correlation function 
   $C(t,\vp)\equiv \langle \theta\left(t,\vp\right)\theta\left(0,-\vp \right) \rangle$
  the following result
 \begin{equation}
C\left(t,\vp\right) = \frac{\epsilon_\theta \epsilon^{-1/3}}{ p^{11/3}} \left\{
\begin{array}{l l}
 C_{\rm s}  \exp\left({-\alpha_{\rm s} \frac{L^2}{\tau_0^2} p^2 t^2}\right)\,,\quad &t\ll \tau_0 \\ 
 C_{\rm \ell} \exp\left({-\alpha_{\rm \ell} \frac{L^2}{\tau_0} p^2 |t|}\right)\,, \quad  &t\gg \tau_0 
 \end{array}
 \right.
\label{eq:2pt-correlation}
\end{equation}
where $\epsilon$ and $\epsilon_\theta$
are the  energy dissipation rates of the velocity and scalar fields respectively,
 and $\tau_0\equiv \left(L^2/\epsilon \right)^{-1/3}$ denotes the eddy-turnover time at the 
energy injection scale,
we refer to \cite{paganiSpatiotemporalCorrelationFunctions2021}
for details. 
The constants $C_{\rm s,\ell}$ and $\alpha_{\rm s,\ell}$ are not universal.
Remarkably, the scalar field inherits in this range
  the temporal properties of the NS velocity field, 
  which exhibits a fully analogous behavior \cite{canetSpatiotemporalVelocityvelocityCorrelation2017, tarpinBreakingScaleInvariance2018}.
  The short time regime is known to be related to the random sweeping effect, while 
  the long time regime was not identified before, but can also be simply interpreted 
    on phenomenological grounds (see below). 
 For NS velocity, the short time Gaussian decay of the two-point and three-point correlation functions was accurately confirmed by DNS \cite{canetSpatiotemporalVelocityvelocityCorrelation2017,gorbunovaSpatiotemporalCorrelations3D2021}.  
 However, the large time exponential regime
 has remained elusive so far in the numerical simulations. 
 Nevertheless, it can be accessed for the scalar field in the synthetic flow
  by tailoring the velocity covariance, as we show in the following.

For Kraichnan's model,
due to the white-in-time nature of the velocity covariance,
the FRG leads to an exponential time decay of the scalar
 correlation function for all times $t$
 as 
 \begin{equation}
 C_{\rm{K}}\left(t,\vp\right) = F\left(p\right) e^{-\kappa_{\rm ren} p^2 |t|}
 \label{eq:2pt-function-kraichnan}
 \end{equation}
   where $F\left(p\right)$ is a complicated  integral.
 In the inertial range, one finds $F\left(p\right) \sim p^{-d-2+\varepsilon}$
while in the weakly non-linear regime (i.e., when the convective term is perturbative), 
$F\left(p\right) \sim p^{-d-2-\varepsilon}$ \cite{paganiSpatiotemporalCorrelationFunctions2021}.
 Moreover, one obtains an explicit expression for the renormalized diffusivity
\begin{eqnarray}
\kappa_{\rm ren} = \kappa +
\frac{d-1}{2d}
\int_{\vp} \frac{D_{0}}{\left(p^{2}+m^{2}\right)^{\frac{d}{2}+\frac{\varepsilon}{2}}}\,.
\label{eq:def-renormalized-mol-visc}
\end{eqnarray}
 Remarkably, the temporal dependence in Eq.~(\ref{eq:2pt-function-kraichnan})
can be derived via several approaches, 
both standard resummation of the self-energy diagrams and FRG calculation \cite{paganiSpatiotemporalCorrelationFunctions2021}. 
An analogous expression was obtained
in~\cite{mitraDynamicsPassiveScalarTurbulence2005,sankarrayUniversalityDynamicMultiscaling2008}
by deriving a differential equation for the two-point correlation function.
 
The FRG approach allows one to go beyond the strict white-in-time limit of Kraichnan's model.
 One finds that
 as soon as the covariance (\ref{eq:vel-2pt-correlation-function}) 
  deviates from pure white noise,
 the two-point correlation function $C_{\rm{K}}\left(t,\vp\right)$ 
  also develops a short time Gaussian regime, 
  which is studied numerically in the following.
  
  An intuitive interpretation of the short- and large-time regimes of the Eulerian spatio-temporal  correlation function can be drawn from the single particle turbulent dispersion, as shown in Ref. 
	\cite{gorbunovaSpatiotemporalCorrelations3D2021}. In this picture, the arguments of the exponentials in  Eq.~(\ref{eq:2pt-correlation}) are related to the mean square displacement of a fluid particle, which
	is proportional to $t^2$ (ballistic regime) at times at which the 
	velocity field is correlated, while at large times, when the velocity is uncorrelated, 
	 it grows linearly in time \cite{taylorDiffusionContinuousMovements1922}, with a coefficient proportional to the eddy diffusivity.

\paragraph{Direct numerical simulations (DNS).} 
To study the  spatio-temporal correlation
 function of a passive scalar, we perform DNS of
  3D homogeneous isotropic turbulent flows. The equations (\ref{eq:scalar-dyn-1}) and (\ref{eq:NS-eq}) are solved on
  a discrete cubic grid with the use of a pseudo-spectral method \cite{canutoSpectralMethodsEvolution2007}.
 Further details on the DNS, 
   snapshots of the velocity and scalar fields and their spatial spectra are provided 
  in \cite{supplMaterial}.
 Once the stationary state is reached, the two-point correlation function $C(t,\vp)$ is calculated
 in spectral domain and averaged over time and
  wavenumber shells as described in  \cite{supplMaterial}.
\begin{figure}
	\centering
	\includegraphics[width=0.99\linewidth]{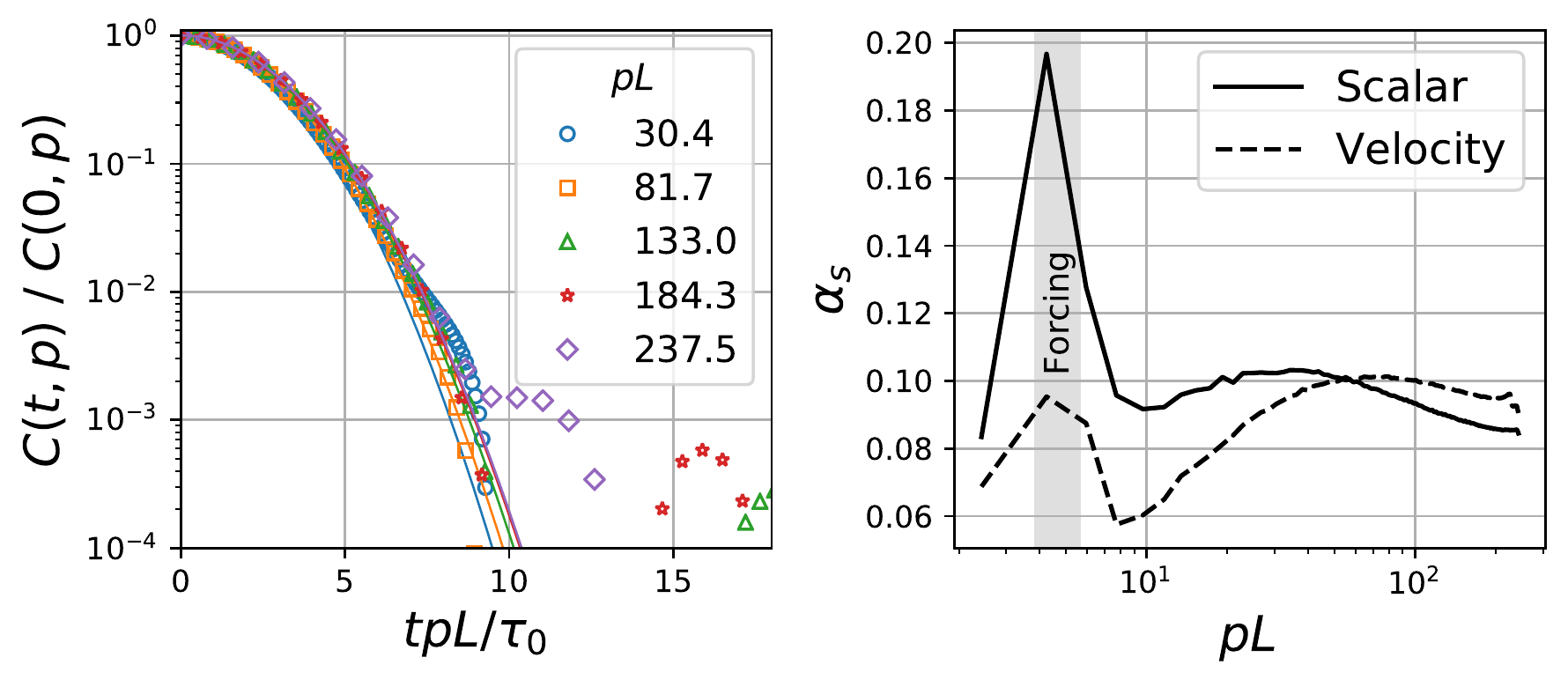}
	\caption{\textit{Left panel:} Normalized averaged two-point correlation function $C(t,\vp)$ of the scalar field in
	the NS velocity field at various wavenumbers $p$, points are the numerical data and plain lines 
	their Gaussian fits.
	All curves collapse when plotted against $tp$ variable. \textit{Right panel:}
	 Dependence on the wavenumber $p$ of the numerically estimated parameter $\alpha_{\rm s}$ for scalar and velocity.
    The Taylor Reynolds is number $R_\lambda = 90$, the Schmidt number $Sc=0.7$.
	In all figures, the wavenumbers are 
   non-dimensionalized by the integral scale $L$,
   and the times by $\tau_0 = L/U_{\textrm{rms}}$ with $U_{\textrm{rms}}$ the root mean square
   velocity.}
	\label{fig:scalarCorrNs}
\end{figure}

\paragraph{Scalar fields advected by NS flows.}
For scalar fields in the inertial-convective range, we expect different behaviors
at short and large time delays according to Eq.~(\ref{eq:2pt-correlation}).
 In the short time regime, 
 we find that the correlation functions obtained from the DNS
 all behave as a Gaussian.
  Their time dependence at fixed wavenumbers $p$ is fitted with 
  the two-parameter function $g(t) = C_0 \exp[-a t^2]$. 
    As illustrated in Fig.~\ref{fig:scalarCorrNs} (left panel), all curves collapse onto a single Gaussian
  when plotted as a function of the variable $tp$, as predicted by
   Eq.~(\ref{eq:2pt-correlation}). 
The value of the fitting parameter $a$ provides a numerical estimation of the decorrelation parameter $\alpha_{\rm s}$ in  Eq.~(\ref{eq:2pt-correlation}) as $\alpha_{\rm s} = a \left(\tau_0/ L p \right)^2 $. 
The dependence of the estimated parameter $\alpha_{\rm s}$ on the wavenumber is displayed in the right panel of Fig.~\ref{fig:scalarCorrNs}, showing that both velocity and scalar reach a plateau beyond the forcing range, 
in agreement with the short-time expression in Eq.~(\ref{eq:2pt-correlation}). 
Moreover, the numerical values of $\alpha_{\rm s}$
for the scalar and velocity fields are very close, which is also in agreement with the FRG result, predicting that $\alpha_{\rm s}$ is solely determined by the properties of the carrier flow and is the same for the velocity and for the scalar. 
Besides, this result demonstrates that the typical decorrelation time scales as $a^{-1/2} \sim p^{-1}$, thus, the Eulerian spatio-temporal correlations are dominated by the sweeping effect.

\par At large time delays $t$, 
one expects a crossover to an exponential decay in time
according to Eq.~(\ref{eq:2pt-correlation}).     
In the DNS, the correlations become very small and oscillatory after the Gaussian decay, preventing the direct 
observation of this crossover.
However, it  can be uncovered in a synthetic flow, as we now discuss.

\paragraph{Scalar fields in Kraichnan's model.}

We perform several sets of simulations, 
where the velocity field is artificially generated and tailored to 
 have a Gaussian statistics. The velocity  covariance is
\begin{equation}
	\left< \hat{v}_i(t_0 + t, \vec{p}) \hat{v}_j^*(t_0, \vec{p}) \right> = \cfrac{D_0}{T_e} 
	\left(p^2+m^2\right)^{\frac{-3-\varepsilon}{2}} P_{ij}(\vp)
	\label{eq:numerical-covariance-Kraichnan-vv}
\end{equation}
for $0<t<T_e$ and vanishes otherwise, with
  $T_e= n\Delta t$, where $\Delta 
t$ is the simulation time step and $n$  the number of 
iterations before the velocity is updated.
 The velocity field is generated in 
Fourier space and fulfills the prescribed spatial covariance, as well 
as the conditions of isotropy and zero divergence. 
For $T_e$ small compared to the dynamical time scales of the flow
($\tau_A \sim \Delta x/U_{\textrm{rms}}$ for advection, and  $\tau_\kappa \sim (\Delta x)^2/\kappa$ for diffusion,
 with  $\Delta x$ spatial grid cell), the velocity field can be considered as white-in-time.
  
\par 
 We analyze 4 sets of simulations allowing to describe
 the various regimes 
 by changing the H\"older exponent and the amplitude of the velocity covariance, as well as the scalar diffusivity. 
Set 1 corresponds to the inertial regime, set 2 to the weakly non-linear regime, sets 3 and 4 to transitional regimes. 
 All parameters are provided in \cite{supplMaterial}. 
We first note that the equal-time two-point scalar correlation functions 
  show a remarkable agreement with the 
 expected power laws  
 for both the inertial and the weakly non-linear regime \cite{supplMaterial}. 
\begin{figure}
	\centering
	\includegraphics[width=0.99\linewidth]{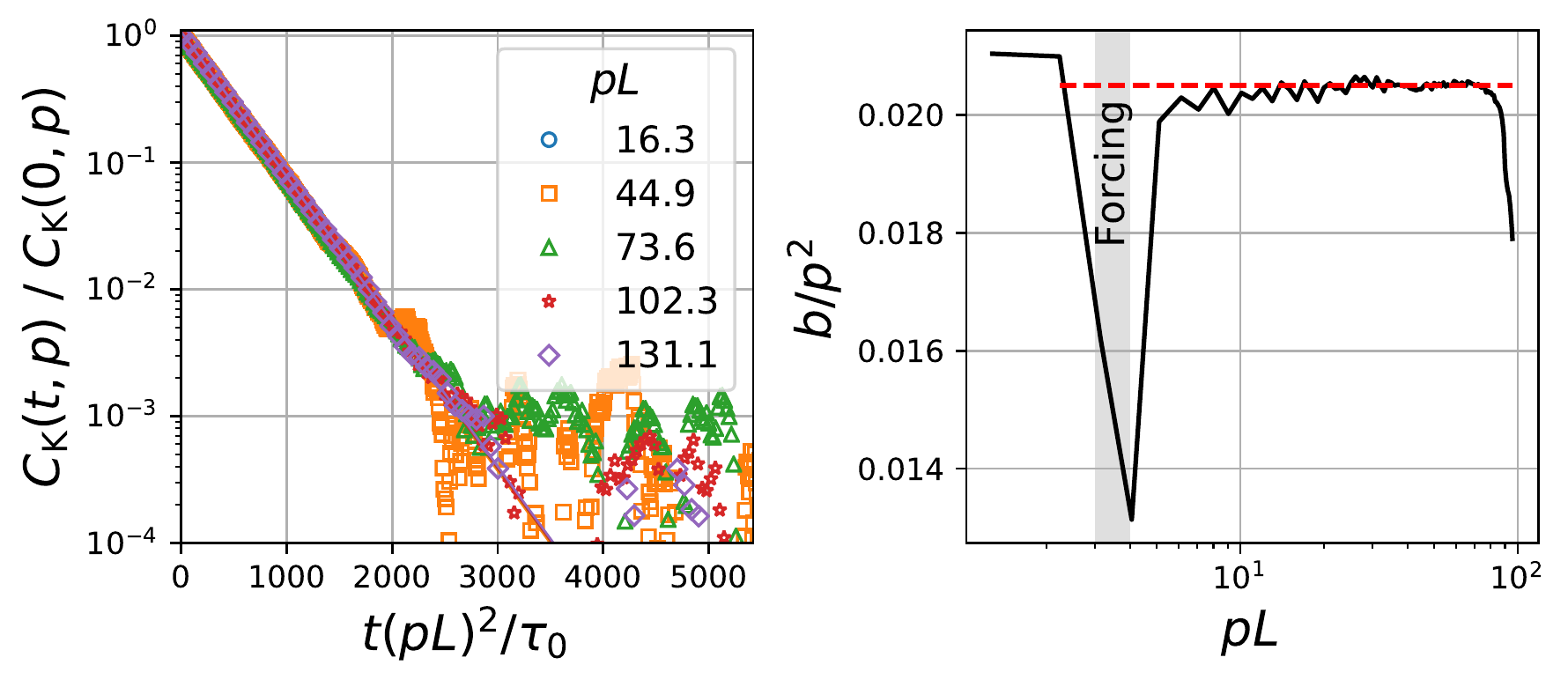}
	\caption{\textit{Left panel:} Two-point correlation function $C_{\rm{K}}(t,\vp)$ at various 
		wavenumbers $p$ of a passive scalar in 3D DNS of Kraichnan's random advection 
		model for $\varepsilon = 1$. All curves collapse onto a single exponential
   		when plotted against the variable $t p^2$.
   		\textit{Right panel}: dependence of the ratio $b/p^2$ on the wavenumber. The dashed line corresponds to the plateau value.}
	\label{fig:scalartimecorrkraichnan}
\end{figure}

 At unequal times, the two-point spatio-temporal correlation function was measured 
numerically for all sets.
 We find that the time dependence of $C_{\rm{K}}(t,\vp)/C_{\rm{K}}(0,\vp)$
 at fixed wavenumbers is always an exponential, as illustrated in Fig.~\ref{fig:scalartimecorrkraichnan} (left panel). 
Moreover, when plotted as a function of $tp^2$,
 all the curves collapse onto a single exponential. 
  Furthermore, we fitted all curves with the two-parameter function $g_K(t)=C_0 \exp[-b t]$. The right panel of  
  Fig.~\ref{fig:scalartimecorrkraichnan} shows that the fitting parameter $b$ compensated by $p^2$ takes an approximately constant value in a large
  range of wavenumbers,
  demonstrating that the correlation functions $C_{\rm{K}}(t,\vp)$ indeed take the expected form (\ref{eq:2pt-function-kraichnan}).
  
In the exponential, the prefactor $\kappa_{\rm ren}$ 
depends on the details of the velocity statistics and cutoffs.
In order to make a precision test of its expression, 
   we varied the amplitude $D_0$, the diffusivity $\kappa$, and H\"older exponent through our 4 sets.

  The plateau value of the ratio  $b/p^2$ gives a numerical estimate of the 
renormalized diffusivity $\kappa_{\rm ren}$ in Eq.~(\ref{eq:2pt-function-kraichnan}). 
\begin{figure}
	\centering
	\includegraphics[width=0.75\linewidth]{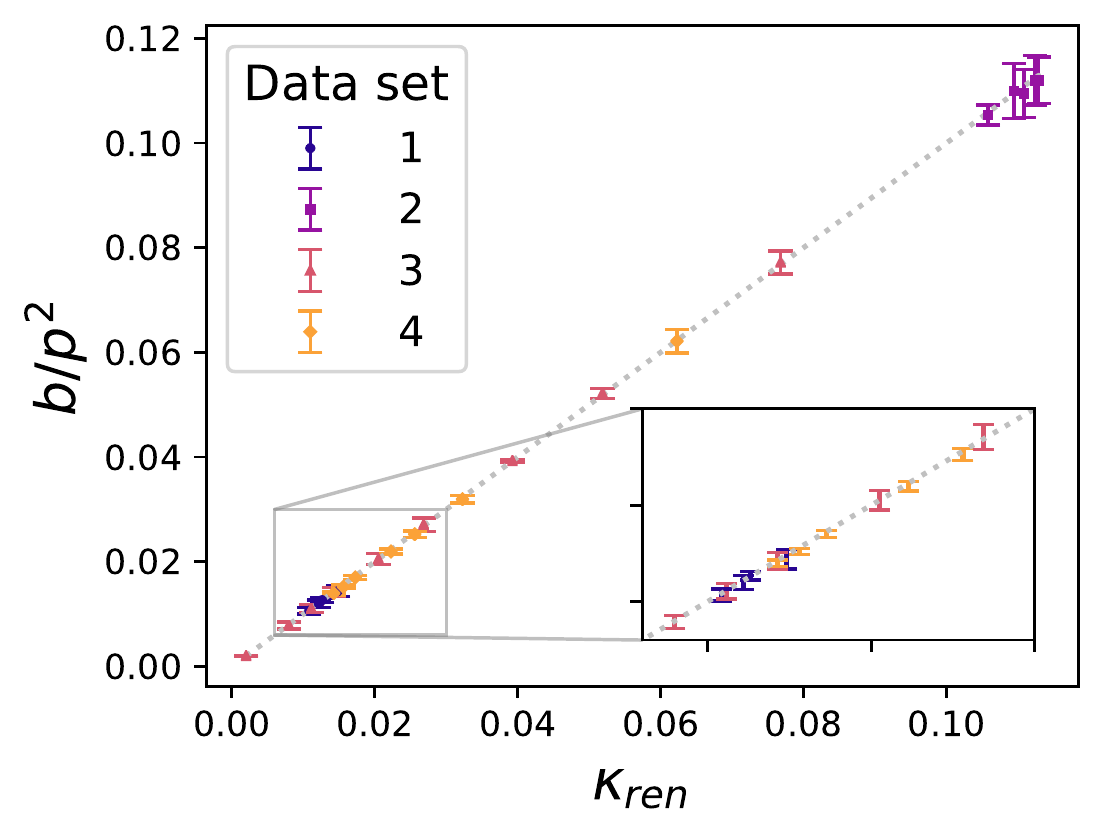}
	\caption{Renormalized scalar diffusivity $\kappa_{\rm ren}$
	  obtained as the decorrelation parameter $b/p^2$ of the exponential fit of the scalar 
		correlation function versus its  theoretical estimate based on the Eq.~(\ref{eq:def-renormalized-mol-visc})
		 for our 4 data sets (see \citep{supplMaterial} for detailed parameters).
		 The error bars correspond to doubled standard deviation of $b/p^2$ from the estimated plateau value, as shown in the right panel of Fig. \ref{fig:scalartimecorrkraichnan}.}
	\label{fig:scalarvsepsilon}
\end{figure}
Besides, the expression for $\kappa_{\rm ren}$ given by
 Eq.~(\ref{eq:def-renormalized-mol-visc})	 
   can be estimated in the simulations upon 
 replacing the integral over wavenumbers by a sum over discrete modes. 
 The value of ($\kappa_{\rm ren} - \kappa$)
 is fixed by the velocity properties only, and can be computed without any 
 information about the scalar. In Fig.~\ref{fig:scalarvsepsilon}, we show the comparison between the 
  value of $\kappa_{\rm ren}$ computed via the correlation fit and via the theoretical expression.
   We obtain a remarkable agreement for all four data sets, 
   thus demonstrating
 that the temporal dependence in Eq.~(\ref{eq:2pt-function-kraichnan})
 is valid for all regimes, even though the equal-time part $F(p)$ behaves differently.
 Let us emphasize that the obtained results are tested for values of $\varepsilon$ spanning up to 1.5, well beyond the perturbative regime \cite{Adzhemyan1998}.
 This analysis hence provides a thorough confirmation of the theory,
    up to the precise form of the prefactor in the exponential.
 
\begin{figure}
	\centering
	\includegraphics[width=0.96\linewidth]{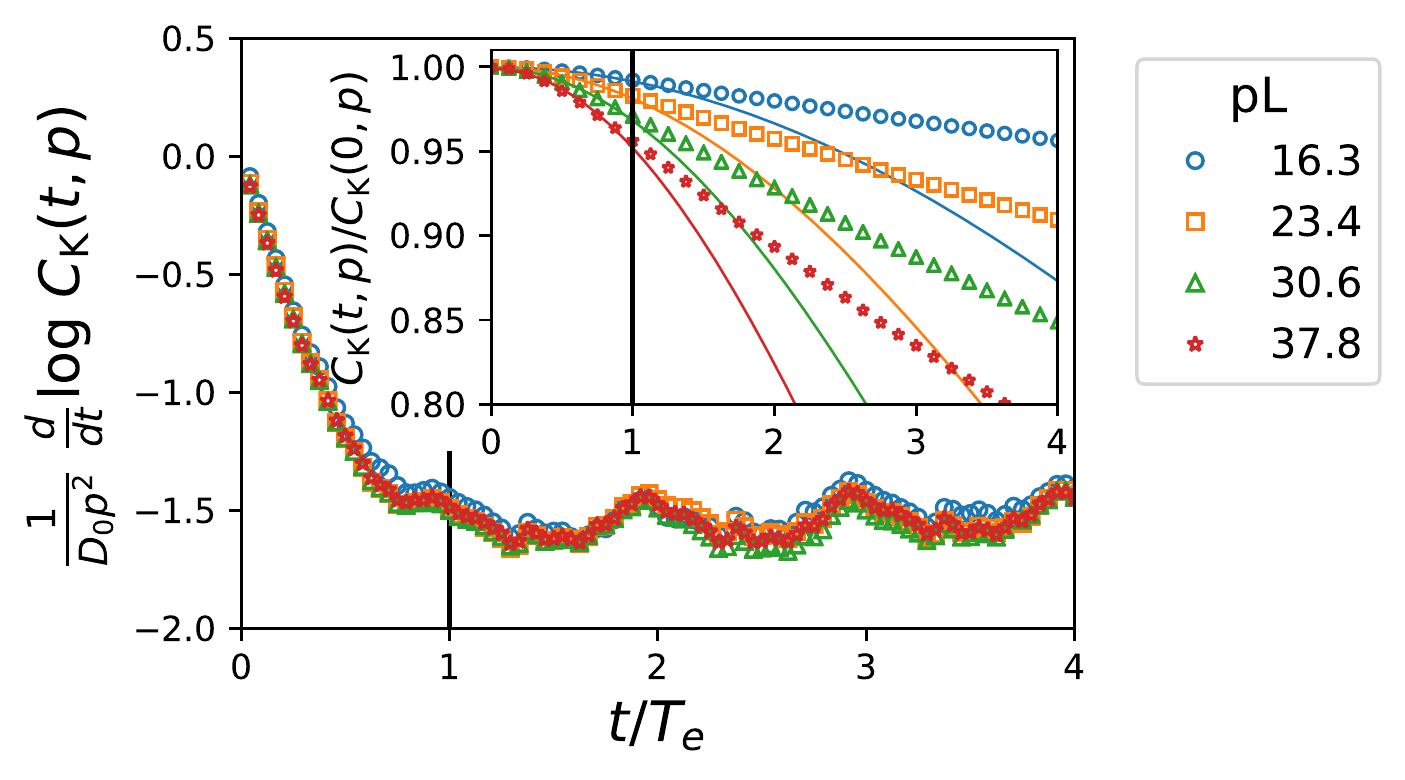}
	\caption{Scalar correlation $C_{\rm{K}}(t,\vp)$ (inset) at time scales comparable to the velocity 
		field renewal period $T_e$ for various wavenumbers $p$. The continuous lines correspond to 
		the Gaussian fit of the points at $t<T_e$.
		 The main panel shows the time derivative of $\log C_{\rm{K}}(t,\vp)$. The linear
		part at $t<T_e$ corresponds to the Gaussian time decay, while the nearly constant part at
		 $t>T_e$ corresponds to an 
		exponential time dependence. The curves at various wavenumbers are rescaled vertically by $p^2$
		 which leads to their collapse.}
	\label{fig:scalarcrossover}
\end{figure}

\paragraph{Crossover between the short- and large-time regimes.} 
 We perform a simulation with a larger renewal time period $T_e$ of the 
velocity field, during  which the scalar is evolving in a 
``frozen'' velocity field. Clearly, the white-in-time approximation
 breaks down in this case.
 The temporal correlations of the velocity field alters
 the small-time behavior of the scalar, which is then expected
 to decay as a Gaussian in $tp$, similarly to the scalar in the NS flow.
 We show in Fig.~\ref{fig:scalarcrossover} (inset) the
correlation function of a passive scalar in a random velocity field updated 
every $n=24$ iterations. 
We found that at times $t<T_e$ its correlation function $C_{\rm{K}}(t,\vp)$ behaves as 
a Gaussian, while at $t>T_e$, the correlation curves 
significantly deviate from the Gaussian and endow an exponential form, 
similar to the ones in Fig.~\ref{fig:scalartimecorrkraichnan}. 
 The transition between the small- and large-time decays
 can be visualized by analyzing the time derivative of $\log C_{\rm{K}}(t,\vp)$. 
When  $C_{\rm{K}}(t,\vp)$ is a Gaussian, 
  this derivative decreases linearly, 
 whereas when  $C_{\rm{K}}(t,\vp)$ is an exponential, it is a negative constant. 
In Fig.~\ref{fig:scalarcrossover}, we show that the derivative of $\log C_{\rm{K}}(t,\vp)$ exhibits the expected 
    crossover from a linear decay to a constant. In addition,
    the rescaling of the derivatives by $1/p^2$ leads to a collapse of all curves, demonstrating that
the Gaussian and the exponential both possess a $p^2$-dependence on the wavenumber, 
 as expected from Eq.~(\ref{eq:2pt-correlation}). 

At high wavenumbers, the large-time regime becomes indiscernible as the scalar field decorrelates fast (because of the $\sim p^2$ dependence) down to near-zero values. We believe that a similar effect hinders the large-time regime in NS flows.

\paragraph{Summary and future prospects.}
Our work shows that the two-point Eulerian spatio-temporal correlation function of 
a passive scalar in a 3D homogeneous isotropic turbulent takes 
 a Gaussian form in the variable $pt$
  at short time delays $t$.
  In addition, the coefficient characterizing the Gaussian time decay of the scalar field is approximately the same as the one of the velocity field.
  According to the FRG prediction, the two-point correlation function possesses a large-time regime, as displayed in Eq.~(\ref{eq:2pt-correlation}). 
  In order to study the crossover between the two regimes, we analyzed the time correlations of a scalar advected by a synthetic velocity field.
We observe that in the case of white-in-time velocity covariance, the correlation function
 accurately follows the expected exponential decay in 
$p^2 t$,  including the precise form of the prefactor of the exponential.
  Moreover, switching from a white-in-time to a non-trivial time covariance of the synthetic velocity field 
 allows one to reveal the crossover between the two time regimes. 
  These results thus confirm the predictions of the FRG analysis
and provide hints that a similar crossover should be present also in the case of scalar fields advected
by NS flows, although it could not be detected in our simulations. 
 It would be very interesting to extend this analysis to higher-order correlation functions. We believe
 that the case of non-passive scalars could also be studied using similar methods.

\smallskip

%%%%%%%%%%%%%%%%%%%%%%%%%%%%%%%%%%%%%%%%%%%%%%%%%%%%%%%%%%
\begin{acknowledgments}
	\par 
	This work was supported by ANR-18-CE92-0019 Grant NeqFluids.
	L.C. and G.B. acknowledge support from Institut Universitaire de France. 
	The simulations were performed using the high performance computing resources from GENCI-IDRIS
	 (grant 020611) and the GRICAD infrastructure (https://gricad.univ-grenoble-alpes.fr), which is supported by Grenoble research communities.
\end{acknowledgments}

\smallskip

\bibliographystyle{apsrev4-1}
%\bibliography{biblio}
\providecommand{\noopsort}[1]{}\providecommand{\singleletter}[1]{#1}%

\pagebreak

\begin{widetext}

\begin{center}
{\bf \Large Appendix}
\end{center}

In this appendix, we provide the detailed description of the simulations performed to obtain the data
 analyzed in the main text, as well as additional figures of the velocity and scalar fields and their spatial spectra.

\section{Simulations of Scalar Advection in Navier-Stokes velocity field}

The direct numerical simulations (DNS) of 
  the Navier-Stokes equation for the turbulent velocity field and advection-diffusion equation for the
   passive scalar are performed in a three-dimensional periodic cubic domain of
  size $2 \pi$, with different resolutions, Reynolds and P\'eclet numbers (see Table \ref{tab:paramsNS}).
   The equations are solved on
  a discrete grid with the use of a pseudo-spectral method  in space 
  and the second order Runge-Kutta scheme for time advancement \cite{canutoSpectralMethodsEvolution2007}. 
  The spatial resolution of the velocity is determined by the condition $p_{max} \eta = 1.5$, where $p_{max}$ is the maximal wavenumber in the computational domain, $\eta$ is the Kolmogorov length scale.
Dealiasing errors are reduced by the polyhedral
  truncation method \cite{orszagNumericalSimulationIncompressible1971}. In order to reach a statistically stationary state of the flow,
  both the velocity and the passive scalar fields are forced randomly in spectral space at large scales \cite{alveliusRandomForcingThreedimensional1999}.
  The random forcing field is independent from the velocity and scalar and is
  updated at each time step of the simulation.
  This forcing scheme yields a numerical approximation of
the stochastic forcing, which is also assumed to be white-in-time. 
An example of instantaneous 
  2D snapshots of the velocity and scalar fields is displayed in the Fig.~\ref{fig:scalarcorrnsfields},
   for the spatial resolution  $N^3 = 256^3$, the Taylor-scale Reynolds number  $R_\lambda = 90$ and the Schmidt number  $Sc = 0.7$. We show in Fig.~\ref{fig:spectrarealscalar} the corresponding kinetic energy spectrum
 of the Navier-Stokes velocity field and the spectrum of variance of the scalar field. The spectra are averaged in time once  the stationary state is reached. 

\begin{table}[h]
	\centering
	\begin{ruledtabular}
	\begin{tabular}{cccccccc}
		\(R_{\lambda}\) & $N_v$ & \(Sc\) & $N_s$ & $Pe$ & \(\Delta t/ \tau_0\) & \(\Delta T_w/\tau_0\) & \(N_w\) \\
		\hline
		60 & 128 & 0.7 & 128 & 222 & $1.4 \times 10^{-3}$ & 0.7 & 795 \\
		60 & 128 & 16 & 512 & 5088 & $3.5 \times 10^{-5}$ & 0.07 & 21 \\
		60 & 128 & 36 & 768 & 11448 & $2.1 \times 10^{-5}$ & 0.03 & 16 \\
		90 & 256 & 0.7 & 256 & 931 & $1.1 \times 10^{-4}$ & 2.94 & 50 \\
	\end{tabular}
	\end{ruledtabular}
	\caption{Parameters of DNS of scalar advection in Navier-Stokes velocity field. $R_\lambda$ - Taylor-scale Reynolds number, $N_s$ - spatial grid resolution for velocity, $Sc$ - scalar Schmidt number, $N_s$ - spatial grid resolution for scalar, $Pe$ - P\'eclet number, $\tau_0$ - large-scale eddy-turnover time, $\Delta t$ - simulation time step, $\Delta T_w$ - width of a time window of correlation measurement, $N_w$ - number of recorded time windows for time correlation.} 
	\label{tab:paramsNS}
\end{table}   
   
\begin{figure}[h]
	\centering
	\includegraphics[width=0.75\linewidth]{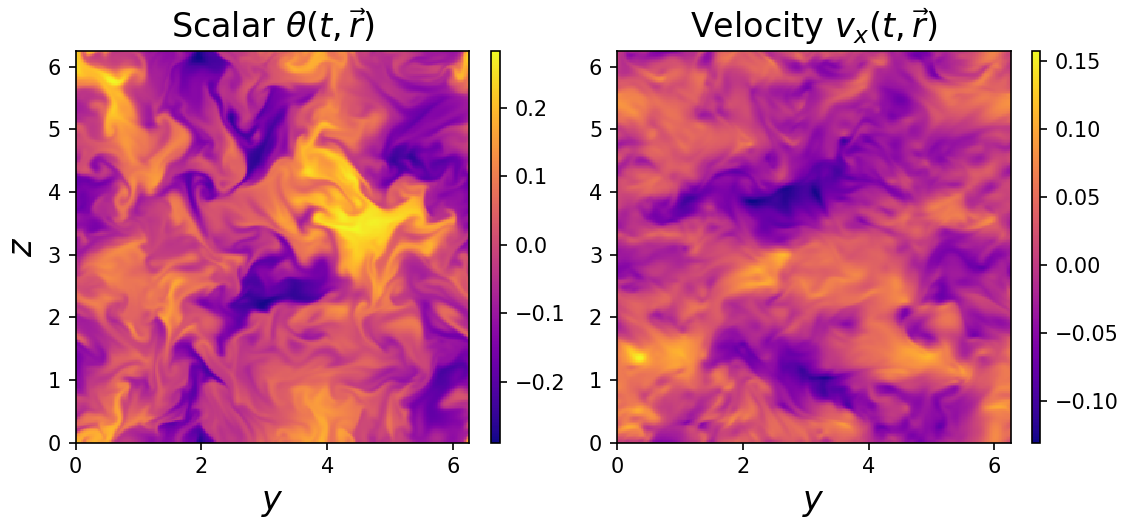}
	\caption{Instantaneous 2D snapshots of the scalar field and of the $x$-component of the velocity field,
	 obtained from the DNS of the 
		Navier-Stokes equation 	at Taylor-scale Reynolds number $R_\lambda = 90$ on a grid 
	of size $N^3=256^3$, with the Schmidt number of the scalar $Sc=0.7$.}
	\label{fig:scalarcorrnsfields}
\end{figure}

\begin{figure}[h]
	\centering
	\includegraphics[width=0.75\linewidth]{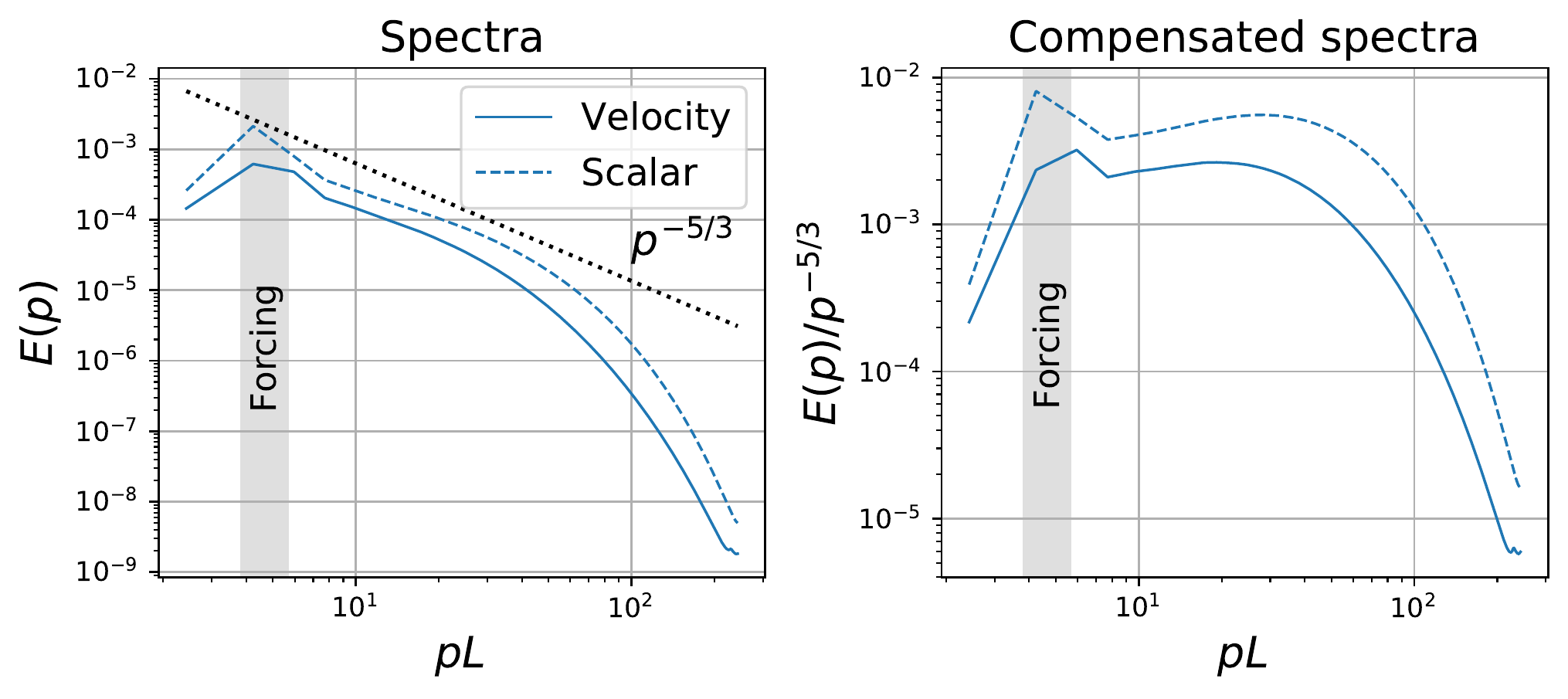}
	\caption{Spectra and compensated spectra of the velocity kinetic energy and scalar variance
	from DNS at Taylor-scale Reynolds number $R_\lambda = 90$ on a grid 
	of size $N^3=256^3$, with the Schmidt number of the scalar $Sc=0.7$.}
	\label{fig:spectrarealscalar}
\end{figure}

The correlation function of the scalar field is also computed in the stationary state, and is averaged in time
 and in space, over spherical spectral shells $S_n$ of thickness $\Delta p =1$, which can be written as:
\begin{equation}
C (t,\vec{p}) = \frac{1}{N_t} \sum_{j=1}^{N_t} \frac{1}{M_n}  \sum_{\vec{p} \in S_n}
 {\mathrm{Re}} \left[ \theta({t_0}_j, \vec{p}) \theta^*({t_0}_j + t, \vec{p}) \right] \,,
\label{eq:def2pcorrNumerical}
\end{equation}
where \(N_t\) is the number of time windows in the simulation, 
and \(M_n\) is the number of 
modes in the spectral spherical shell \(S_n\). 
The same computation was 
simultaneously performed for the velocity correlations. 

\section{Simulations of Scalar Advection in synthetic velocity field}

To approximate the stochastic white-in-time self-similar velocity field of the
Kraichnan model of passive scalar transport, we implemented a synthetic field generator
which realizes the following variance: 
\begin{eqnarray*}
	\avg{\hat{u}_i(t_0, \vec{p}) \hat{u}_j^*(t_0+t, \vec{p})} =
	\begin{cases}
	    \cfrac{D_0}{T_e} 
    	\left(p^2+m^2\right)^{\frac{-3-\eps}{2}} P_{ij}(\vec{p}) \qquad  &  t< T_e\\
        0   &  t>T_e
	\end{cases}
\end{eqnarray*}
  $T_e$ thus represents the existence time of an individual realization of the velocity field,
  which is  calculated as $T_e = n \Delta t$, where $n$ is  the number of time iterations before a 
new velocity field is generated. The infra-red cut-off wavenumber $m$ is set to 
1 in all simulations. When $T_e$ is negligible compared to the other characteristic
 time scales of the flow, we obtain an approximation of the $\delta$-function in time.

\par The velocity field is generated in Fourier space in the following way:
\begin{equation}
	\hat{\vec{u}}(t,\vec{p}) = \sqrt{\cfrac{2 D_0 (p^2+m^2)^{\frac{-3-\eps}{2}} 
		}{T_{e}}} \vec{\eta}(t, \vec{p})
\end{equation}
where $\vec{\eta}$ is a unit random complex vector fulfilling the requirements of
zero-divergence and isotropy.
 The velocity field can also be characterized by the root-mean-square 
velocity $u_{RMS} = \sqrt{2 E_k /3}$, with $E_k$ the total kinetic energy. It is 
linked with $D_0$ by $u_{RMS} = \sqrt{A \frac{D_0}{2 T_e}}$, where $A = \frac{1}{3} 
\sum_{\vec{p}} \left(p^2+m^2\right)^{\frac{-3-\eps}{2}}$ is the constant coming 
from the numerical estimation of the integral of the spatial velocity spectrum.
 The random velocity field is generated in spectral space every $n$ 
iterations and is totally uncorrelated with the previous realizations.

The scalar field
is also subjected to a large-scale random forcing. The forcing term $f_\theta$ is 
generated at each iteration in spectral space within a narrow wavenumber band $3<p_f<4$
corresponding to large scales. The amplitude of the forcing is the same in all 
simulations.  The dynamics of the scalar can be characterized by  a non-dimensional parameter analogous to the P\'eclet number \cite{falkovichParticlesFieldsFluid2001} $Pe = 
D_0^\prime L^\varepsilon/ \kappa$, where $D_0^\prime = D_0/\eps$, and $L$ 
corresponds to the scalar integral length scale $L=2\pi/p_f$. 
For each simulation the renormalized scalar diffusivity was estimated numerically according to  Eq.~(6)
 of the main text:
\begin{equation}
	\kappa_{\rm ren} = \kappa +
	\frac{1}{3} \sum_p \frac{D_{0}}{\left(p^{2}+m^{2}\right)^{\frac{d}{2}+\frac{\varepsilon}{2}}} = \kappa + A D_0
	\label{eq:kapparen}
\end{equation}

\par The parameters of all the simulations are provided in Table \ref{tab:parametersKraichnan}. 
The simulations are divided into four sets. The sets 1 and 2 
contain simulations at various $\eps$ with the parameter $D_0^\prime$ 
kept constant. 
Low values of the parameter $\eps$ correspond to rough velocity fields, 
while higher $\eps$ values correspond to smoother ones. 
Instantaneous snapshots of 
the velocity fields at various $\eps$ are shown in  Fig. 
\ref{fig:velocitykraichnanfields}. In the set 1 the diffusivities $\kappa$ were 
chosen small ($Pe > 1$) to reach the inertial regime, dominated by the advection. 
In the set 2 the diffusivity is fixed at a higher value, so that $Pe \sim 0.1$. In 
this case the scalar transport is dominated by diffusion but still affected by the 
advection. For this reason we refer to this regime as weakly non-linear. 
The snapshots of the scalar fields for the data sets 1 and 2 are provided in 
Figs.~\ref{fig:scalarkraichnanfields} and \ref{fig:scalarkraichnanfieldsdissipative}, 
respectively.

\begin{table}[h]
	\begin{ruledtabular}
	\begin{tabular}{c|ccccccccccc}	
		set &$N$ & $\kappa$ & $\varepsilon$ & $D_0$ & $n$ & $\Delta t$ & $U_{RMS}$ & $ Pe $ & $A$ & $N_w$ & $\kappa_{\rm ren}$ \\
		\hline
		\multirow{4}{1em}{1} & 256 & $6\times 10^{-3}$ & 0.1 & $4.18\times10^{-4}$ & 6 & $0.24\times10^{-3}$ & 2.87 & 1.45 & 14.56 & 8 & $1.22\times10^{-2}$  \\
		& 256 & $1.70\times 10^{-3}$ & 0.5 & $2.09\times10^{-3}$ & 6 & $0.24\times10^{-3}$ & 4.22 & 6.54 & 6.28 & 10 & $1.48\times10^{-2}$ \\
		& 384 & $1.7\times 10^{-4}$ & 1.0 & $4.18\times10^{-3}$ & 9 & $0.16\times10^{-3}$ & 3.54 & 85.7 & 2.96 & 17 & $1.25\times10^{-2}$ \\
		& 384 &$2.5\times 10^{-5}$ & 1.5 &  $6.27\times10^{-3}$ & 9 & $0.16\times10^{-3}$ & 3.32 & 793 & 1.72 & 17 & $1.08\times10^{-2}$ \\
		\hline
		\multirow{4}{1em}{2} & 192 & 0.1 & 0.1 & $4.18\times10^{-4}$ & 6 & $0.32\times10^{-3}$ & 2.42 & 0.09 & 13.81 & 26 & $10.58\times10^{-2}$  \\
		& 192 & 0.1 & 0.5 & $2.09\times10^{-3}$ & 6 & $0.32\times10^{-3}$ & 3.62 & 0.11 & 6.16 & 5 & $11.29\times10^{-2}$ \\
		& 192 & 0.1 & 1.0 & $4.18\times10^{-3}$ & 6 & $0.32\times10^{-3}$ & 3.55 & 0.15 & 2.96 & 5 & $11.23\times10^{-2}$ \\
		& 192 & 0.1 & 1.5 &  $6.27\times10^{-3}$ & 6 & $0.32\times10^{-3}$ & 3.32 & 0.20 & 1.72 & 5 & $11.08\times10^{-2}$ \\
		\hline
		\multirow{9}{1em}{3} & 64 & $2\times 10^{-3}$ & 1.0 & 0.0 & - & $0.98\times10^{-3}$ & 0.0  & 0.0 & -& 440 & $2\times10^{-3}$ \\
		& 64 & $2\times 10^{-3}$ & 1.0 & $2.09\times10^{-3}$ & 2 & $0.98\times10^{-3}$ & 2.86 & 2.5 & 2.86 & 440 & $0.80\times10^{-2}$ \\
		& 96 & $2\times 10^{-3}$ & 1.0 & $3.14\times10^{-3}$ & 3 & $0.65\times10^{-3}$ & 3.05 & 3.76 & 2.91 & 318 & $1.11\times10^{-2}$ \\
		& 128 & $2\times 10^{-3}$ & 1.0 & $4.18\times10^{-3}$ & 4 & $0.49\times10^{-3}$ & 3.23 &  5.02& 2.93 & 266 & $1.42\times10^{-2}$ \\
		& 192 & $2\times 10^{-3}$ & 1.0 & $6.28\times10^{-3}$ & 6 & $0.32\times10^{-3}$ & 4.34 &  7.52& 2.95 & 95 & $2.05\times10^{-2}$ \\
		& 256 & $2\times 10^{-3}$ & 1.0 & $8.37\times10^{-3}$ & 8 & $0.24\times10^{-3}$ & 5.03 & 10.04 & 2.97 & 49 & $2.68\times10^{-2}$ \\
		& 384 & $2\times 10^{-3}$ & 1.0 & $12.55\times10^{-3}$ & 12 & $0.16\times10^{-3}$ & 6.16 & 22.52 & 2.97 & 35 & $3.92\times10^{-2}$ \\
		& 512 & $2\times 10^{-3}$ & 1.0 & $16.75\times10^{-3}$ & 16 & $0.12\times10^{-3}$ & 7.12 & 30.03 & 2.98 & 35 & $5.18\times10^{-2}$ \\
		& 768 & $2\times 10^{-3}$ & 1.0 & $25.10\times10^{-3}$ & 32 & $0.06\times10^{-3}$ & 8.74 & 45.05 & 2.98 & 39 & $7.68\times10^{-2}$ \\
		\hline
		\multirow{7}{0.5em}{4} & \multirow{7}{1.5em}{128} & 0.05 & \multirow{7}{1em}{1.0} & \multirow{7}{5em}{$8.36\times10^{-3}$} & \multirow{7}{0.5em}{4} & \multirow{7}{5em}{$0.49\times10^{-3}$} & \multirow{7}{2em}{3.23} & 0.30 & \multirow{7}{2em}{2.93} & \multirow{7}{1em}{34} & $6.22\times10^{-2}$ \\
		&& 0.02 &&&&&& 0.75 & & & $3.22\times10^{-2}$ \\
		&& 0.013 &&&&&& 1.12 & & & $2.56\times10^{-2}$ \\
		&& 0.01 &&&&&& 1.50 & & & $2.22\times10^{-2}$ \\
		&& 0.005 &&&&&& 3.00 & & & $1.72\times10^{-2}$ \\
		&& 0.0033 &&&&&& 4.50 & & & $1.56\times10^{-2}$ \\ 
		&& 0.002 &&&&&&	7.51 & & & $1.43\times10^{-2}$ \\
	\end{tabular}
	\end{ruledtabular}
	\caption{Parameters of simulations of scalars in synthetic velocity fields. $N$ - spatial resolution of the 
		computational grid, $\kappa$ - scalar diffusivity, $\eps$ - H\"older exponent of the velocity field,
		 $D_0$ - amplitude of the velocity covariance, $n$ - number of iterations between velocity field updates, 
		 $\Delta t$ - simulation time step, $U_{RMS}$ - root-mean-square velocity, $Pe$ - 
		P\'eclet number, $A$ - sum of the velocity spectrum, $N_w$ - 
		numbers of time windows for correlations averaging, $\kappa_{\rm ren}$ - 
		numerical estimation of the renormalized diffusivity estimated according 
		to  Eq.~(\ref{eq:kapparen})}	
	\label{tab:parametersKraichnan}
\end{table}

We also generated two additional sets of data in order to test
the dependence of $\kappa_{\rm ren}$ as given by Eq.~\ref{eq:kapparen} 
on the amplitude $D_0$ of the velocity covariance  
 and on the bare diffusivity $\kappa$. Thus,
the set 3 consists of simulations at a fixed 
 value of diffusivity $\kappa$ and $\eps = 1$, with gradually increasing the velocity 
covariance amplitude $D_0$. The first simulation at $D_0 = 0$ corresponds to the purely 
diffusive case. The  set 4 consists of a single simulation in which 7 scalars 
with various diffusivities are transported simultaneously by the same velocity 
field, with $\eps = 1$ and fixed $D_0$.

We show in  the Fig. 
\ref{fig:spectrascalarinertial} the equal-time correlation functions for the velocity and the scalar field
 in the set 1, which is the inertial regime, and in the set 2, which  is the weakly non-linear regime.
 The corresponding power-laws $p^{-d-2+\varepsilon}$ and $p^{-d-2-\varepsilon}$ respectively are accurately observed for all values of $\varepsilon$.

\begin{figure}[h]
	\centering
	\includegraphics[width=0.99\linewidth]{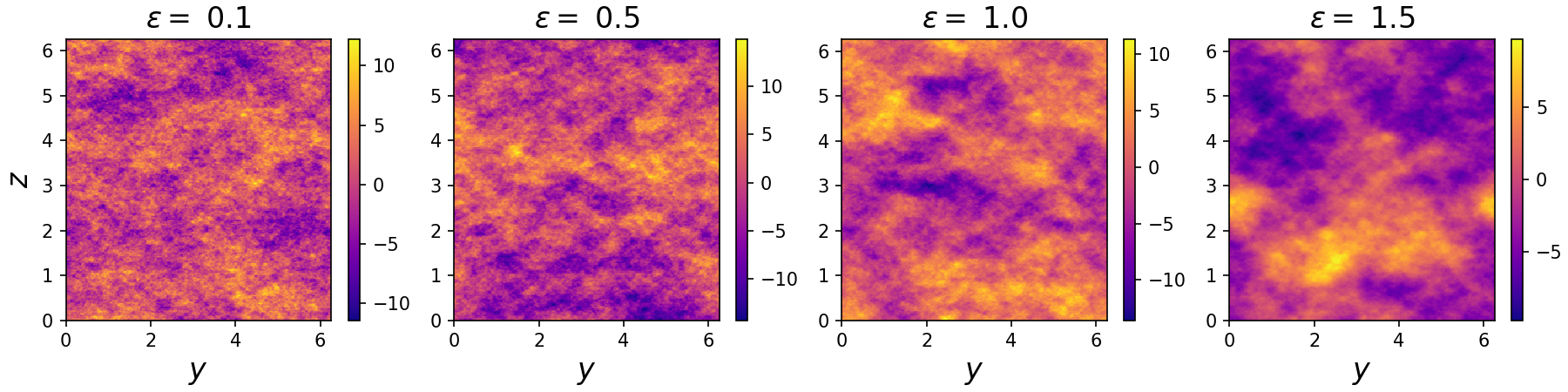}
	\caption{Two-dimensional instantaneous cuts of the $x$-component of the synthetic velocity field at various $\epsilon$.}
	\label{fig:velocitykraichnanfields}
\end{figure}

\begin{figure}[h]
	\centering
	\includegraphics[width=0.99\linewidth]{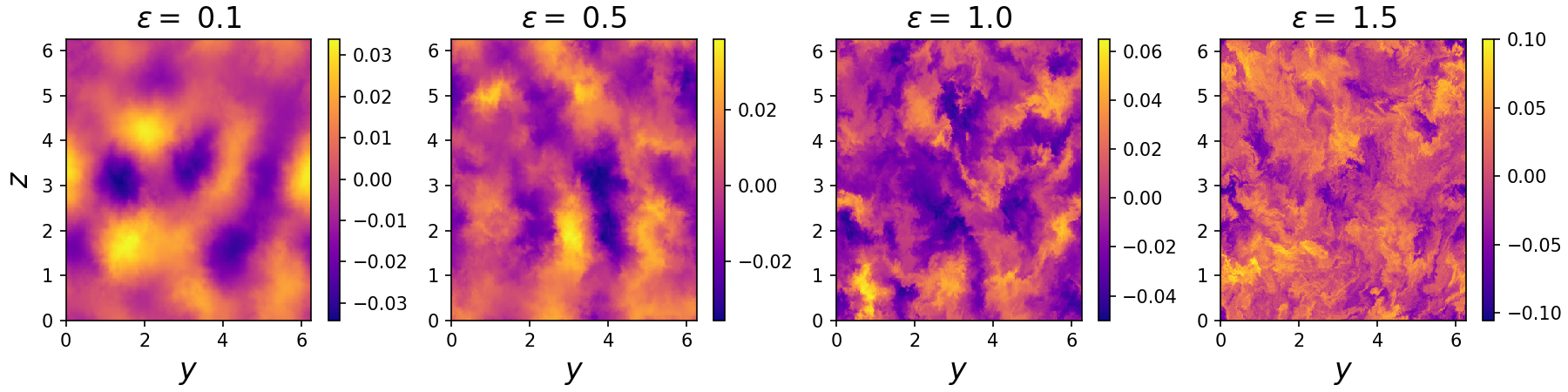}
	\caption{Two-dimensional instantaneous  cuts of the passive scalar field at various $\varepsilon$ in the inertial regime (corresponding to set 1).}
	\label{fig:scalarkraichnanfields}
\end{figure}

\begin{figure}[h]
	\centering
	\includegraphics[width=0.99\linewidth]{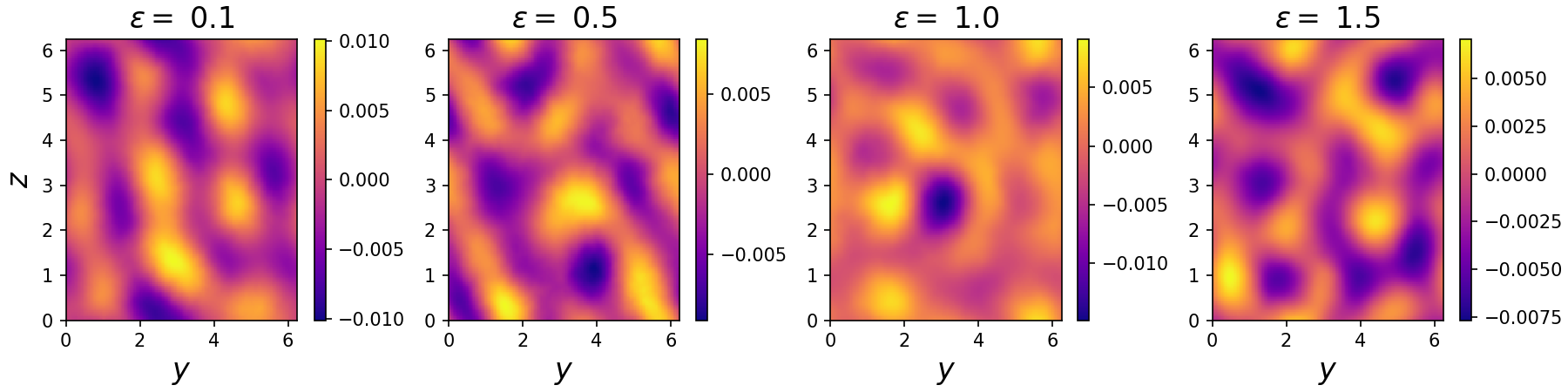}
	\caption{Two-dimensional instantaneous  cuts of the passive scalar field at various $\varepsilon$ in the weakly non-linear regime (corresponding to set 2).}
	\label{fig:scalarkraichnanfieldsdissipative}
\end{figure}

\begin{figure}[h]
	\centering
	\includegraphics[width=0.95\linewidth]{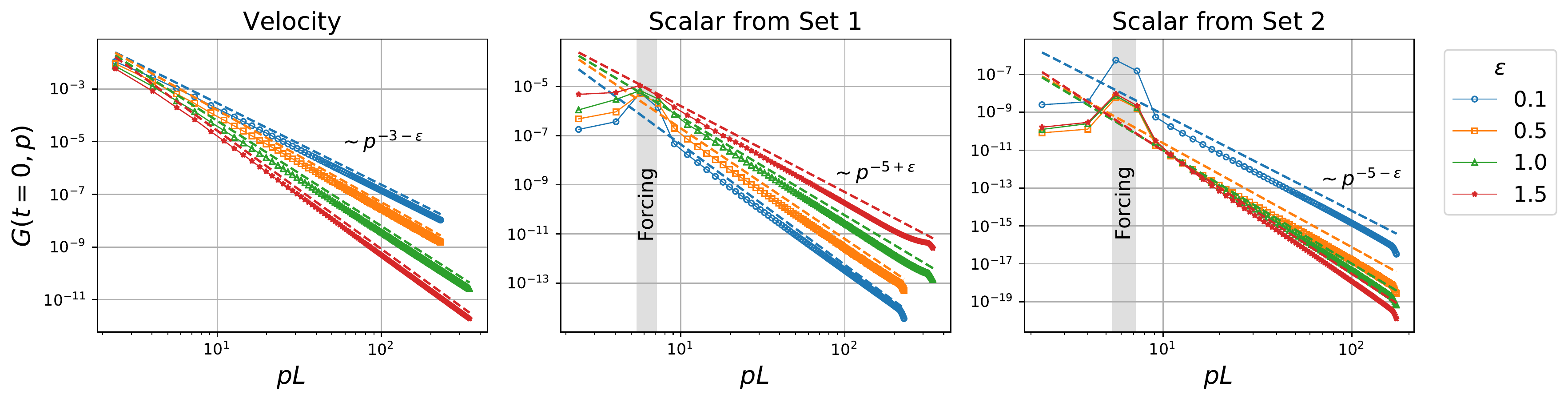}
	\includegraphics[width=0.95\linewidth]{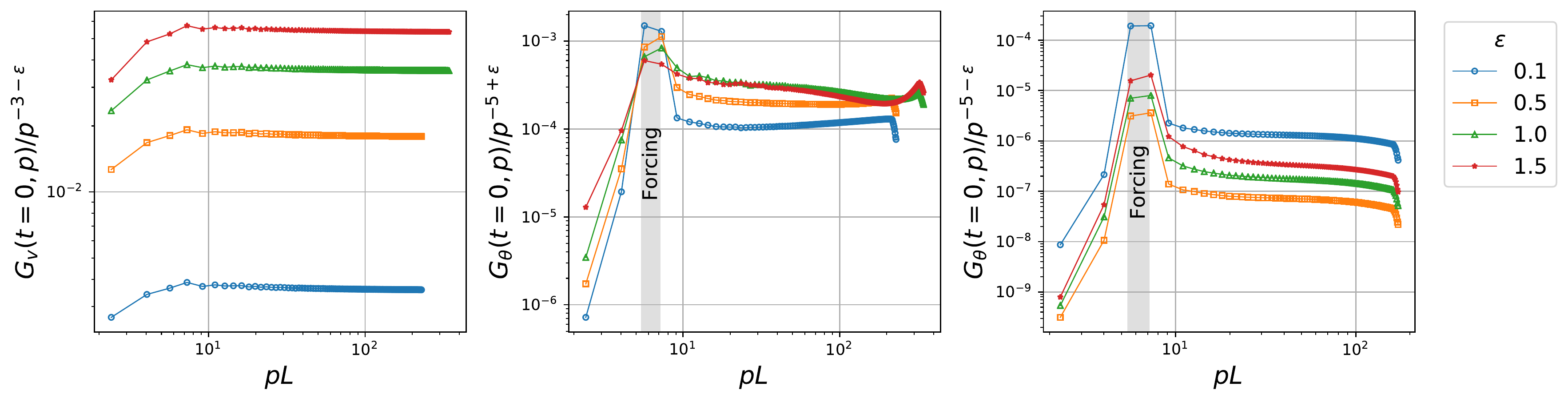}
	\caption{Upper row: equal-time two-point correlation function of the velocity and scalars from data set 1 (inertial 
		regime) and set 2 (weakly non-linear regime). Points correspond to the 
		numerical data, the dashed lines to the corresponding theoretical power 
		law. The bottom row show the same data but compensated by the 
		theoretically predicted power law.}
	\label{fig:spectrascalarinertial}
\end{figure}

\providecommand{\noopsort}[1]{}\providecommand{\singleletter}[1]{#1}%

\end{widetext}

\end{document}